\documentclass[twocolumn,english,aps,prl,twocolumn,superscriptaddress]{revtex4}
\usepackage[T1]{fontenc}
\usepackage[latin9]{inputenc}
\usepackage{textcomp}
\usepackage{amssymb}
\usepackage{amsmath}
\usepackage{graphicx}
\usepackage{esint}
\usepackage{svg}
\usepackage{physics}
\usepackage{tikz}
\usepackage{subfigure}
\usepackage{hyperref}
\usepackage{bbold}
\usepackage{babel}
\usepackage{ulem}
\usepackage{amsmath}

\hypersetup{
colorlinks = true,
linkcolor= teal,
citecolor= teal,
urlcolor= blue
}

\usepackage{bm}
\usepackage{braket}
\makeatletter

\@ifundefined{textcolor}{}
{%
 \definecolor{BLACK}{gray}{0}
 \definecolor{WHITE}{gray}{1}
 \definecolor{RED}{rgb}{1,0,0}
 \definecolor{GREEN}{rgb}{0,1,0}
 \definecolor{BLUE}{rgb}{0,0,1}
 \definecolor{CYAN}{cmyk}{1,0,0,0}
 \definecolor{MAGENTA}{cmyk}{0,1,0,0}
 \definecolor{YELLOW}{cmyk}{0,0,1,0}
 }
\makeatother

\begin{document}

\title{The Power of One Clean Qubit in Supervised Machine Learning}

\author{Mahsa Karimi}
\thanks{mahsa.karimi1@ucalgary.ca}
\affiliation{Department of Physics and Astronomy, University of Calgary, Calgary, AB, T2N 1N4, Canada
}\affiliation{
Institute for Quantum Science and Technology, University of Calgary, Calgary, AB, T2N 1N4, Canada}
\author{Ali Javadi-Abhari}
\affiliation{IBM Quantum, IBM T. J. Watson Research Center, Yorktown Heights, NY, USA}
\author{Christoph Simon}
\affiliation{Department of Physics and Astronomy, University of Calgary, Calgary, AB, T2N 1N4, Canada
}\affiliation{
Institute for Quantum Science and Technology, University of Calgary, Calgary, AB, T2N 1N4, Canada}
\author{Roohollah Ghobadi}
\thanks{rghobadi@ucalgary.ca}

\affiliation{Department of Physics and Astronomy, University of Calgary, Calgary, AB, T2N 1N4, Canada
}\affiliation{
Institute for Quantum Science and Technology, University of Calgary, Calgary, AB, T2N 1N4, Canada}

\date{\today}

\begin{abstract}

This paper explores the potential benefits of quantum coherence and quantum discord in the non-universal quantum computing model called deterministic quantum computing with one qubit (DQC1) in supervised machine learning. We show that the DQC1 model can be leveraged to develop an efficient method for estimating complex kernel functions. We demonstrate a simple relationship between coherence consumption and the kernel function, a crucial element in machine learning. The paper presents an implementation of a binary classification problem on IBM hardware using the DQC1 model and analyzes the impact of quantum coherence and hardware noise. The advantage of our proposal lies in its utilization of quantum discord, which is more resilient to noise than entanglement.

\end{abstract}
\maketitle{}
\subsection{INTRODUCTION}

Recent progress in the control and mitigation of noise and decoherence has paved the way for the development of intermediate-scale quantum devices consisting of hundreds of qubits. Although these devices are currently not fault-tolerant, there is considerable evidence that they possess superior computational capabilities compared to classical supercomputers, as a result of their ability to support quantum entanglement~\cite{zhong2020quantum,madsen2022quantum}. As quantum hardware continues to evolve, it is expected to play a crucial role in various fields such as quantum simulations, quantum chemistry, and quantum machine learning (QML)~\cite{preskill2018quantum,huang2022quantum}.

The use of quantum hardware for complex computations such as kernel function estimation has been proposed as a way to achieve a quantum advantage in machine learning~\cite{schuld2019evaluating,havlivcek2019supervised}. Quantum entanglement is considered a key resource for this~\cite{rebentrost2014quantum, lloyd2014quantum,gao2022enhancing}, but it is highly susceptible to noise, thus it is important to explore other forms of quantum correlation that are less sensitive to noise or require less entanglement.

The Deterministic Quantum Computing with One Qubit (DQC1) model is a non-universal quantum computing model that leverages a single qubit as a probe to interact with a highly mixed quantum state and estimate computationally expensive functions. This ability is known as the ``power of one qubit''~\cite{knill1998power}. The DQC1 model generates quantum discord, a resilient type of weak quantum correlation, using the coherence of a pure qubit~\cite{ollivier2001quantum,modi2012classical}. Quantum discord is more resistant to noise than entanglement and may offer a quantum advantage in noisy conditions for quantum illumination tasks~\cite{weedbrook2016discord}.

There is a limited body of literature exploring the use of DQC1 in machine learning contexts~\cite{park2018noise, ghobadi2019power, vedaie2020quantum}. Reference\cite{park2018noise} investigates the advantages of DQC1 in addressing the parity learning problem. In reference~\cite{ghobadi2019power}, DQC1 is proposed for application in kernel based supervised machine learning. Finally, reference~\cite{vedaie2020quantum} builds upon the results in reference~\cite{ghobadi2019power}, extending the concept to multiple kernel learning for supervised machine learning within a DQC1 framework.

This paper studies the use of the DQC1 model in supervised machine learning for efficient estimation of complex kernel functions. The study is implemented on IBM hardware and examines the effects of coherence consumption, quantum discord, and hardware noise. The DQC1 protocol reduces measurement errors by only measuring one qubit, achieving high classification accuracy despite requiring more gates than a similar protocol in~\cite{havlivcek2019supervised}.

The paper is structured as follows: Section~\ref{pre} provides a review of the DQC1 algorithm, quantum coherence and quantum discord, and a brief overview of kernel-based supervised machine learning.
In Sec.~\ref{supervised machine learning}, we describe the application of DQC1 for the estimation of arbitrary kernel functions.
Sec.~\ref{implementation} presents our implementation of supervised machine learning using DQC1 on IBM hardware. We also compare the DQC1 kernel with the projected quantum kernel in this section.
Sec.~\ref{role of coh} presents the role of quantum coherence and the effect of noise in our implementation. 
Finally, Sec.~\ref{conclusion} discusses the limitations imposed on the DQC1 kernel and concludes with a summary of our findings.

\section{PRELIMINARIES}\label{pre}
\subsection{DQC1}\label{sec:DQC1}

The DQC1 model was originally introduced in the context of nuclear magnetic resonance (NMR) quantum information processing and has been implemented in various physical setting~\cite{passante2011experimental,lanyon2008experimental, hor2015deterministic,wang2019witnessing}. 

As shown in Fig.\ref{DQC1-circuit}, the DQC1 circuit consists of one control qubit which is prepared in $\frac{I_{1}+\alpha Z}{2}$ with $\alpha\in[0,1]$, $Z$ as Pauli $Z$ matrix and $n$ target qubits in a maximally mixed state denoted by $\frac{I_{n}}{2^{n}}$ where $I_{n}$ is a $2^n\times2^n$ identity matrix. One can change the purity of the control qubit by tuning $\alpha\in[0,1]$: for $\alpha=0$ and $\alpha=1$ the control qubit will be in maximal mixed and pure states, respectively, while for $0<\alpha<1$ the control qubit will be in a partially mixed state. Once the control qubit is evolved through the Hadamard gate, as shown in Fig.\ref{DQC1-circuit}, the matrix form of the initial state in the computational basis of control qubit becomes

\begin{equation}\label{rhoin}
\rho_{in}=\frac{1}{2^{n + 1}}\begin{pmatrix} I_{n} &\alpha I_{n}\\ \alpha I_{n}&I_{n}
\end{pmatrix}.
\end{equation}
Following the application of the DQC1 circuit evolution 
\begin{equation}\label{Udqc1}
U_{DQC1}=|0\rangle\langle 0|\otimes I_{n}+| 1\rangle\langle 1|\otimes U_{n},
\end{equation} 
where $U_{n}$ is an arbitrary $2^n\times2^n$ unitary matrix that is applied to $n$ target qubits, the total state is updated to 
\begin{equation}\label{rhof}
\rho_{f}=U_{DQC1}\rho_{\text{in}}U^{\dagger}_{DQC1}=\frac{1}{2^{n+1}}\begin{pmatrix} I_{n}&\alpha U_{n}^{\dagger}\\ \alpha U_{n}&I_{n}
\end{pmatrix}.
\end{equation}

Tracing out the last $n$ qubits from Eq.(\ref{rhof}), the density matrix of the control qubit denoted by $\rho_{f,c}$ is given by
\begin{equation}\label{rhoc}
\rho_{f,c}=\frac{1}{2}\begin{pmatrix} 1&\frac{\alpha}{2^{n}}\text{tr}(U_{n}^{\dagger})\\ \frac{\alpha}{2^{n}}\text{tr}(U_{n})&1
\end{pmatrix}.
\end{equation}

Measuring the off-diagonal elements of the control qubit can be used to calculate the trace of a unitary matrix, as demonstrated in Eq. (\ref{rhoc}). The number of measurements needed to estimate the off diagonal elements in Eq.(\ref{rhoc}) within precision $\epsilon$ and with probability $1-\delta$ is $O(\epsilon^{-2}\alpha^{-2}\log(1/\delta))$, which is independent of the number of register qubits. Therefore, DQC1 serves as an efficient method for estimating the trace of an arbitrary unitary matrix, a problem for which no efficient classical algorithm is known\cite{knill1998power}. In the special case of a real positive semi-definite matrix, there is a classical randomized algorithm to estimate the trace~\cite{avron2011randomized}. As pointed out in \cite{aaronson2017computational}, the capability of DQC1 to achieve universal classical computation is still an open question. However, there exist complexity arguments that prove the classical efficient approximation of the output probability distribution of the DQC1 model is impossible unless the polynomial-time hierarchy collapses to the second level \cite{fujii2018impossibility}.

For completeness, we will now demonstrate that the clean qubit and register qubits remain in a separable state throughout the computation as stated in~\cite{poulin2004exponential}. To do this, we will use the eigenvectors and eigenvalues of $U_n$, denoted as $|u_i\rangle$ and $\lambda_i$, respectively. In the basis of $\{|u_i\rangle\}$, the mixed state of the target qubits can be represented as $I_{n}=\sum_{i}|u_i\rangle \langle u_i|$. Applying the DQC1 evolution as defined in Eq.(\ref{Udqc1}) results in

\begin{equation}
U_{DQC1}(|0\rangle+|1\rangle)|u_{i}\rangle=(|0\rangle+\lambda_{i}|1\rangle)|u_{i}\rangle,
\end{equation}
which is clearly a product state, meaning there is no entanglement between the control qubit and the target qubits.

\subsection{B.DQC1 resource}

The DQC1 model has been widely studied to investigate the potential use of quantum resources, other than entanglement, in quantum computation~\cite{datta2005entanglement,datta2008quantum}. In this section, we shortly review the definitions of quantum coherence and discord and we demonstrate that it is the consumption of this coherence that allows for the production of discord~\cite{ma2016converting}. 

The rigorous definition of coherence was first given in~\cite{baumgratz2014quantifying}.
The coherence is defined as

\begin{equation}\label{cohdef}
C(\rho)=S(\rho_{\text{diag}})-S(\rho),
\end{equation}
where $S(\rho)=-\text{tr}(\rho\text{log}\rho)$ is the von Neumann entropy and $\rho_{\text{diag}}$ is the diagonal part of $\rho$. One can find the change of coherence for the control qubit defined as $\Delta C=C(\rho_{in,c})-C(\rho_{f,c})$ with $\rho_{in,c}$ and $\rho_{f,c}$ as input and output state for the control qubit, respectively. The input state of the control qubit can be obtained by tracing out the Eq.~\ref{rhoin} (See Eq.S1 in supplementary material). Using Eq. S1 and Eq.\ref{rhoc} in Eq.(\ref{cohdef}) one obtains~\cite{park2018noise}

\begin{equation}\label{deltac}
\Delta C = H_{2}\left(\frac{{1-\alpha\frac{|\text{tr}(U_{n})|}{2^n}}}{2}\right) - H_{2}\left(\frac{1 -\alpha}{2}\right),
\end{equation}
where $H_{2}(x)=-x\text{log}_{2}x-(1-x)\text{log}_{2}(1-x)$ is the binary Shannon entropy (see supplementary material for the derivation). From Eq.(\ref{deltac}), it is clear that the coherence consumption which is determined by the parameter $\alpha$ and the trace of $U_{n}$, can be obtained efficiently by DQC1. 

Quantum discord is a generalization of the classical notion of mutual information and is defined as the difference between the total quantum mutual information and the classical mutual information of the subsystems. 
For a bipartite system in state $\rho_{AB}$, the quantum discord is defined by the difference
\begin{equation}
D(\rho_{AB})=I(\rho_{AB})-J(\rho_{AB}),
\end{equation}
with $I(\rho_{AB})$ and $J(\rho_{AB})$ as the quantum mutual information, and the measurement-based mutual information, respectively. The quantum mutual information is given by
\begin{equation}
I(\rho_{AB})=S(\rho_{A})+S(\rho_{B})-S(\rho_{AB}).
\end{equation}
The measurement-based mutual information on the other hand is given by
\begin{equation}
J(\rho_{AB})=S(\rho_{B})-\text{min}_{E_{k}}[p_{k}S(\rho_{B|k})],
\end{equation}
where the minimum is taken over all possible positive operator-valued measurements (POVM) $\{E_{k}\}$ on subsystem $A$, $\rho_{B|k}=\text{tr}_{A}(E_{k}\rho_{AB})/p_{k}$ is the post-measurement state for system $B$ if outcome $k$ is obtained with probability $p_{k}=\text{tr}(E_{k}\rho_{AB})$.

In the following, we use an alternative definition for discord known as geometric discord, which is easier to calculate and has a closed form for DQC1 \cite{passante2012measuring}. For a given quantum state $\rho$, the geometric discord is defined as \cite{dakic2010necessary}

\begin{equation}\label{geodis}
D_{G}(\rho)=\text{min}_{\chi\in\mathcal{C}}||\rho-\chi||^{2},
\end{equation}
where $\mathcal{C}$ denotes the set of classical zero-discord states and $||A-B||^{2}=\text{tr}(A-B)^{2}$. Evaluating Eq.(\ref{geodis}) for the state in Eq.(\ref{rhof}) leads to \cite{passante2012measuring}

\begin{equation}\label{disdqc1}
D_{G}(\rho_{f})=(\frac{\alpha}{2})^{2}\frac{1}{2^{n}}\left(1-\frac{|\text{tr}(U_{n}^{2})|}{2^{n}}\right).   
\end{equation}

\begin{figure}
\centering
\includegraphics[width=0.4\textwidth]{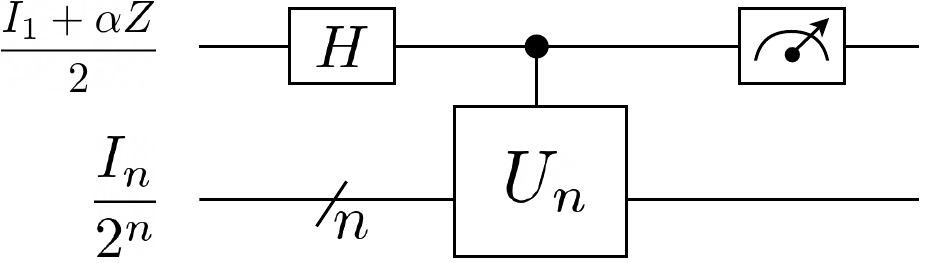}
\caption{The circuit representation of the DQC1 algorithm. The input states for control and target qubits are $\frac{I_{1}+\alpha Z}{2}$, with $\alpha\in[0,1]$ and $\frac{I_{n}}{2^{n}}$, respectively. $H$ and $Z$ denote the Hadamard and Pauli $Z$ gates, respectively.} 
\label{DQC1-circuit} 
\end{figure}

Evaluating Eq.(\ref{disdqc1}) can be done by applying two consecutive controlled-$U_n$ in Fig. \ref{DQC1-circuit}. In \cite{ma2016converting}, the connection between coherence consumption and discord production in DQC1 was examined, where it was demonstrated that quantum discord is bounded by the quantum coherence consumed in the control qubit, i.e.

\begin{equation}\label{disbound}
D_{G}(\rho_{f})\leq\Delta C.
\end{equation}

In our implementation, the results obtained from IBM hardware verifies the relation  (\ref{disbound}).

\subsection{SUPERVISED MACHINE LEARNING: SUPPORT VECTOR MACHINES AND KERNEL METHOD}
\begin{figure}
\centering
\includegraphics[width=0.4\textwidth]{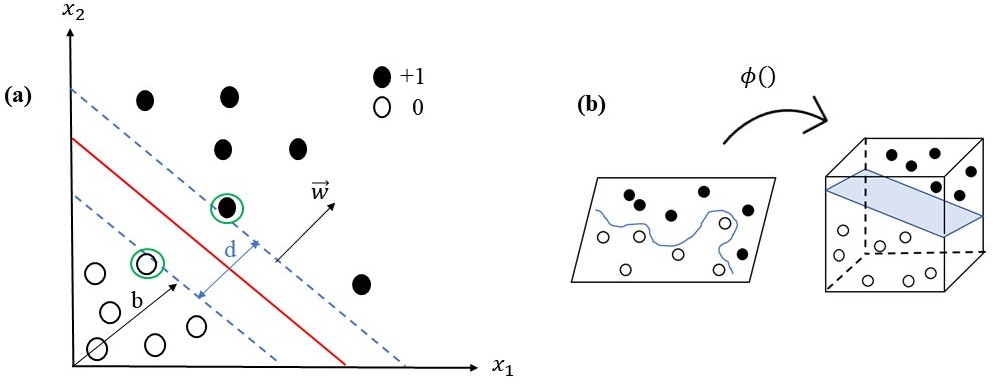}
\caption{~\textbf{(a)} A Support Vector Machine (SVM) is a classifier used to separate two linearly separable classes, depicted in black and white. The data points closest to the decision boundary (shown in red), one from each class, are known as support vectors and are indicated by green circles.
~\textbf{(b)} When data points of two classes cannot be separated by a hyperplane in the original space (left), a non-linear mapping can be applied to project the data points into a higher-dimensional feature space (right) where a hyperplane can be found to separate the classes.}
\label{linear}
\end{figure}

In this section, we introduce the concepts of support vector machines (SVM) and the kernel method within the context of supervised machine learning. Given a set of $n$ training data points, represented by $X_{\text{train}}:=\{(x_i,y_i): i=1,2,...,n\}$, where each data point has $k$ features, i.e., $x_{i}\in\mathbb{R}^{k}$, and is labeled by $y_i\in\{1,-1\}$. The task is to use the training data to develop a classifier function that can accurately predict the labels for test (unseen) data. In the simplest scenario, where the data points are linearly separable, the classifier function can be expressed as

\begin{equation}
f(x) =\text{sign}\big(w^T x + b\big),
\end{equation}
where $w\in \mathbb R^k$ and $b$ are to be determined such that $y_{i}f(x_i)>0$. In the SVM, the separating plane $f(x)$ is determined by maximizing the distance between the hyperplane to the nearest data point of each class ~\cite{tong2001support}, see Fig. \ref{linear}(a).

The SVM can be generalized to the case of non-linearly separated data points by mapping the data points to a higher dimension space for which the data is linearly separable, see Fig.~\ref{linear}(b). In other words, one considers a non-linear mapping $\phi:X\to\mathcal{H}$, so that the decision function can be written as 

\begin{equation}
f(x) = \text{sign}\big(w^T \phi(x)+ b\big).
\end{equation}

In this context, $\mathcal{H}$ and $\phi(x)$ are known as feature space and feature map, respectively. 

It is well-known that for nonlinear separable data points the SVM leads to solutions of the form~\cite{hofmann2008kernel},

\begin{equation}\label{fkernel}
f(x)= \text{sign}\sum_{i}{\beta^{*}_{i}K(x,x_{i})},
\end{equation}
where $\beta^{*}_{i}$ are coefficients to be determined, and we defined the kernel function $K(x,x_{i})=\langle\phi(x),\phi(x_{i})\rangle$, where $\langle,\rangle$ denotes the inner product in feature space $\mathcal{H}$. The procedure for finding $\beta_{i}$ in Eq.(\ref{fkernel}) is through maximizing

\begin{equation}\label{lagrange}
\sum_{i=1}^{n}\beta_{i}-\frac{1}{2}\sum_{i,j=1}^{n}y_{i}y_{j}\beta_{i}\beta_{j}K(x_{i},x_{j}),
\end{equation} 
over the training data, subject to $\sum_{=1}^{n}\beta_{i}y_{i}=0$ and $\beta_{i}\geq0$. For a positive definite kernel, Eq.(\ref{lagrange}) is a concave problem, whose solution $\beta^{*}=(\beta_{1}^{*},...,\beta_{n}^{*})$ can be found efficiently.

The basic idea of SVM can be extended to the quantum domain by interpreting the feature map as a quantum state that can be constructed by a quantum circuit, and the kernel function as the inner product between respective quantum states~\cite{schuld2019quantum,havlivcek2019supervised}.

\section{SUPERVISED MACHINE LEARNING WITH DQC1}\label{supervised machine learning}
 
The freedom in choosing the unitary operator $U_{n}$ in DQC1 allows one to make a connection between DQC1 and the kernel method~\cite{ghobadi2019power}. To see this, we choose $U_{n}=u^l(\vec x)\left(u^l(\vec x')\right)^{\dagger}$, where $u^l$ represents $l$ consecutive application of unitary operator $u$ with $\vec x$ and $\vec x'$ as encoded data points in the gate parameters. Next, we note that $\text{tr}(u^l(\vec x)\left(u^l(\vec x')\right)^{\dagger})$ is positive semidefinite, i.e. $\sum_{i,j}c_{i}c_{j}\text{tr}(u^l(\vec x)\left(u^l(\vec x')\right)^{\dagger})\geq0$ for $c_{i}\in\mathbb{C}$.

Rewriting Eq.(\ref{rhoc}) for $U_{n}(\vec x,\vec x')=u^l(\vec x)\left(u^l(\vec x')\right)^{\dagger}$,

\begin{equation}\label{rhokernel}
\rho_{f,c} = \frac{1}{2}
\begin{pmatrix}1&\alpha K^{*}(\vec x,\vec x')\\\alpha K(\vec x,\vec x')&1
\end{pmatrix},
\end{equation}
where $K(\vec x,\vec x')=\frac{\text{tr}(U_{n}(\vec x,\vec x'))}{2^{n}}$. 

From equation (\ref{rhokernel}), it follows that the DQC1 model allows for an efficient method for estimating arbitrary, complicated kernel functions.

Interestingly, by comparing equation (\ref{rhokernel}) with equation (\ref{deltac}), we can relate the coherence consumption to the kernel function. For example, by setting $\alpha=1$ in equation (\ref{deltac}), we obtain

\begin{equation}\label{deltac2}
\Delta C(\vec x,\vec x')=H_{2}\left(\frac{1-|K(\vec x,\vec x')|}{2}\right). 
\end{equation}

From Eq.(\ref{deltac2}) the following key insight can be obtained. Firstly, when $\Delta C(\vec x,\vec x')=0$ it follows that $K(\vec x,\vec x')=1$, indicating that the kernel function is incapable of distinguishing between the two data points. This lack of discrimination hinders the learning process and underscores the significance of coherence consumption in learning. Secondly, Eq.(\ref{deltac2}) provides a means to observe the influence of hardware noise. To see this, note that in the absence of hardware noise, one expects that $\Delta C(\vec x,\vec x)=0$ as $K(\vec x,\vec x)=1$. In real situations, where noise cannot be ignored, the diagonal elements of a kernel will be smaller than one, related to a loss of coherence as shown by Eq.(\ref{deltac2}) (See Sec.\ref{role of coh} for more details.). Please note that it follows from Eq.(\ref{deltac}) that the above conclusions are applicable for arbitrary $\alpha\in(0,1]$. 

\section{IMPLEMENTATION ON IBM HARDWARE}\label{implementation}
\begin{figure}
\centering
\includegraphics[width=0.2\textwidth]{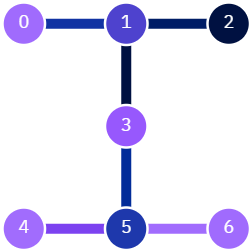}
\caption{Architecture of $7$-qubit ``$ibm\_perth$'' quantum device.}
\label{hardware}
\end{figure}

In this section, we describe our implementation of supervised machine learning based on the DQC1 model. Our scheme was implemented on the ``$ibm\_perth$'' quantum processor, as shown in Fig.\ref{hardware}, using IBM's open-source software interface, Qiskit. 

Our demonstration were performed on ibm\_perth, which is a $7-$qubit superconducting quantum processor from IBM. It consists of fixed-frequency transmon qubits connected according to the coupling map of Fig.~\ref{hardware}. The device characteristics at the time of our demonstration were as follows:
Median $T1: 137.24 \mu s$
Median $T2: 107.32 \mu s$
Median 1-qubit gate error: $0.03\%$
Median 2-qubit gate error: $1.06\%$
Median readout error: $4.56\%$
The code was implemented on top of Qiskit Machine Learning, which provides the dataset and the SVM kernels. The code and Qiskit version for these demonstration can be found online~\cite{code}.

\begin{figure}
\centering
\includegraphics[width=0.4\textwidth]{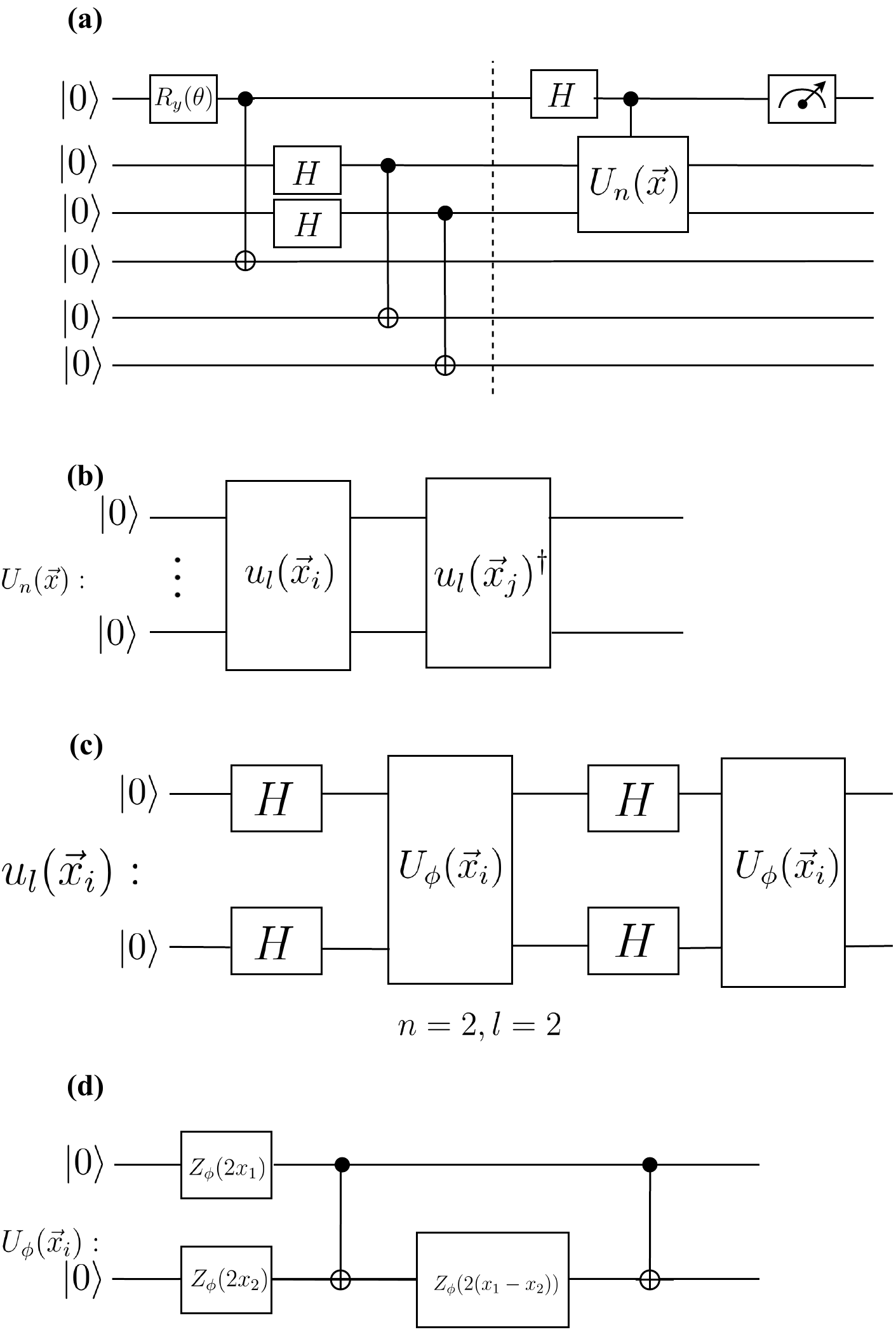}
\caption{\textbf{(a)}~A schematic picture of a three-qubit version of DQC1 circuit, with one control-qubit, and two target qubits and three ancilla qubits. The first part of the circuit, before the dashed line, prepares the control qubit in $\frac{I_{1}+\alpha Z}{2}$ and target qubits in mixed state. Here, $R_{y}(\theta)=\exp (-i\frac{\theta }{2}Y)$, is a rotation gate around $y$ axis, and $Y$ denotes Pauli $Y$ gate. \textbf{(b)}~The gate decomposition for the unitary matrix  \mbox{$U_{n}=u^{l}(\vec x_i)\left(u^{l}(\vec x_j)\right)^\dagger$}. $u^l(\vec x_i)$ is the encoding circuit, where $l$ is the number of the iterations of this gate decomposition (length of the circuit). \textbf{(c)}~ The gate components of the unitary operator $u^l{(\vec{x_i})}$ adapted from~\cite{havlivcek2019supervised} for two target qubits $n=2$, and two iterations $l=2$. Here $U_\phi(\vec{x_i})$ is the feature map (defined in the text). At first, the Hadamard gate is applied to all qubits and then a diagonal gate $U_{\phi}(\vec x_i)$ acts on the qubits. \textbf{(d)}~ The feature encoding circuit $U_{\phi}(\vec x_i)$ in Eq.(\ref{uphi}).   
Here, $Z_\phi=\text{diag}(1,e^{i\phi(\vec{x_i})})$ is a single qubit phase gate.}
\label{dqc}
\end{figure}

Fig.\ref{dqc}~(a) shows the schematics of our implemented circuit which is composed of two target qubits and three ancilla qubits. In the first part of the circuit, left to the dashed line, input states for control and target qubits are prepared. The mixed state preparation of target qubits are based on creating Bell states between target and ancilla qubits, followed by ignoring the state of ancilla qubits. The resulting state of the control and target qubits right before the dashed line is given by $\rho_{\text{control}}\otimes\frac{I_{2}}{4}$ where $\rho_{\text{control}}=\text{diag}(\cos^{2}\frac{\theta}{2},\sin^{2}\frac{\theta}{2})$. By choosing $\theta=2\cos^{-1}(\sqrt{\frac{1+\alpha}{2}})$ the state of control qubit becomes $\frac{I_{1}+\alpha Z}{2}$. 

In our implementation on the IBM real hardware, the purity of the resulting target qubits is $\text{tr}(\rho^{2}_{target})=0.506$, which deviates from the ideal mixed state by $6\times10^{-3}$.

To benchmark the performance of our protocol, we use the encoding map and dataset used in ~\cite{havlivcek2019supervised}. In~\cite{havlivcek2019supervised} the data points $\vec x$ and $\vec x'$ are mapped onto gate parameters of the unitary matrix $U_{2}(\vec x,\vec x')=u^{l}(\vec x)\left(u^l(\vec x')\right)^{\dagger}$, where

\begin{equation} 
u^{l}(\vec x)=\prod\limits_{i = 0}^{l} (U_{\phi(\vec x)}H^{\otimes 2})_{i},
\label{ul}
\end{equation}
with $l$ as the number of iterations of each layer in the feature map (See Fig.\ref{dqc}(b)), encoding map $U_{\phi (\vec x)}$, and $H^{\otimes 2}$ denotes two Hadamard gates acting on two qubits (see Fig.\ref{dqc}(c)).

\begin{equation}\label{uphi}
U_{\phi (\vec x)} = \exp (\sum\limits_{S \subseteq [n]} {{\phi _S}}(\vec x)\prod\limits_{i\in S} {Z_i} ),
\end{equation}
where $\phi_{i}(\vec x)=x_{i}$, $\phi_{i,j}(\vec x)=(\pi-x_{i})(\pi-x_{j})$, and $Z_{i}$ denotes Pauli $Z$ gate (See Fig.\ref{dqc}(d). Fig.~\ref{dqc}(c) shows the quantum circuit that describes Eq.(\ref{ul}) for $l=2$. We defined our kernel as $K(\vec x,\vec x')=\frac{|\text{tr}(U_2(\vec x,\vec x'))|}{4}$. It has been conjectured that approximation of the resulting kernel function for the encoding map Eq.~(\ref{uphi}) with $l=2$ is hard classically, i.e. the resources
required to perform it, increase at a non-polynomial rate with respect to number of qubits \cite{havlivcek2019supervised}.

Our implementation is divided into three phases. In the first stage, we run the circuit in Fig.~\ref{dqc} for all pairs of training data to obtain the corresponding density matrix for the control qubit, using the quantum state tomography package in Qiskit with repeating each measurement $8000$ times (shots), and therefore to obtain the corresponding kernel function. Having obtained the kernel function on the quantum hardware, we apply the classical SVM to obtain the optimal separating hyperplane, or equivalently $\beta^{*}$ by applying Eq.(\ref{lagrange}). Finally, in the prediction phase, given test data $\vec x$, we run the DQC1 circuit to estimate the $K(\vec x,\vec x_{i})$ for all $\vec x_{i}\in X_{\text{train}}$ and apply Eq.(\ref{fkernel}).\\

\begin{figure}
\centering
\includegraphics[width=0.5\textwidth]{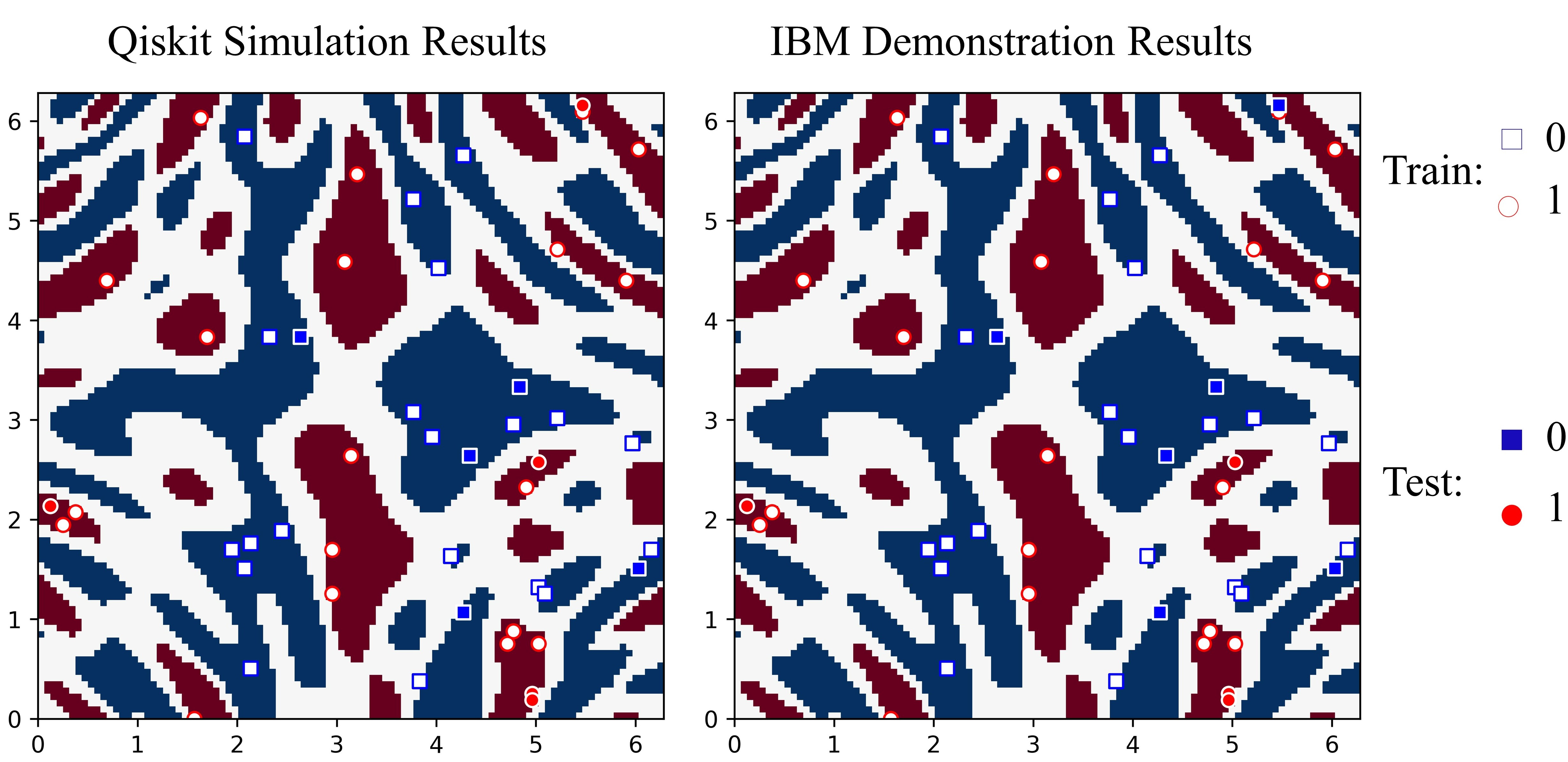}
\caption{The Qiskit simulation (left) and IBM (right) results for the DQC1 kernel classification are shown in Fig.\ref{dqc} with $n=2$ and $l=2$. We used the ``ad\_hoc" dataset, which includes $20$ training and $5$ test samples per label. The accuracy of classification for the IBM quantum simulator (Qiskit) is $100\%$, while it is $90\%$ for IBM's real hardware.}
\label{adhoc-classification}
\end{figure}

In Fig.~\ref{adhoc-classification}, we display the results of applying the above procedure for the classification task on the ``ad\_hoc'' dataset for the IBM simulator (Qiskit) (left) and IBM 7-qubit hardware (right) for $l=2$ for the control qubit in the pure state, i.e. $\alpha = 1$ \cite{code}. From Fig. (\ref{adhoc-classification}), it can be seen that the accuracy of the Qiskit simulator is $100\%$. On the other hand, the obtained accuracy on the hardware is $90\%$. The difference between the simulation and hardware performance can be attributed to the effects of hardware noise. It is worth noting that the circuits were optimized using the Qiskit compiler, specifically the Approximate Quantum Compilation method ~\cite{madden2022best}. This method converts the entire circuit (excluding ancilla) to a 3-qubit unitary matrix, and then re-synthesizes it into a new circuit that approximates the matrix with 0.995 accuracy (synthesis fidelity). A higher synthesis fidelity uses more CNOT gates in the resulting circuit, which reduces approximation error but increases runtime noise. This method reduced the CNOT gate count of the circuit from $177$ to $19$.\\

\begin{figure}
\centering
\includegraphics[width=0.4\textwidth]{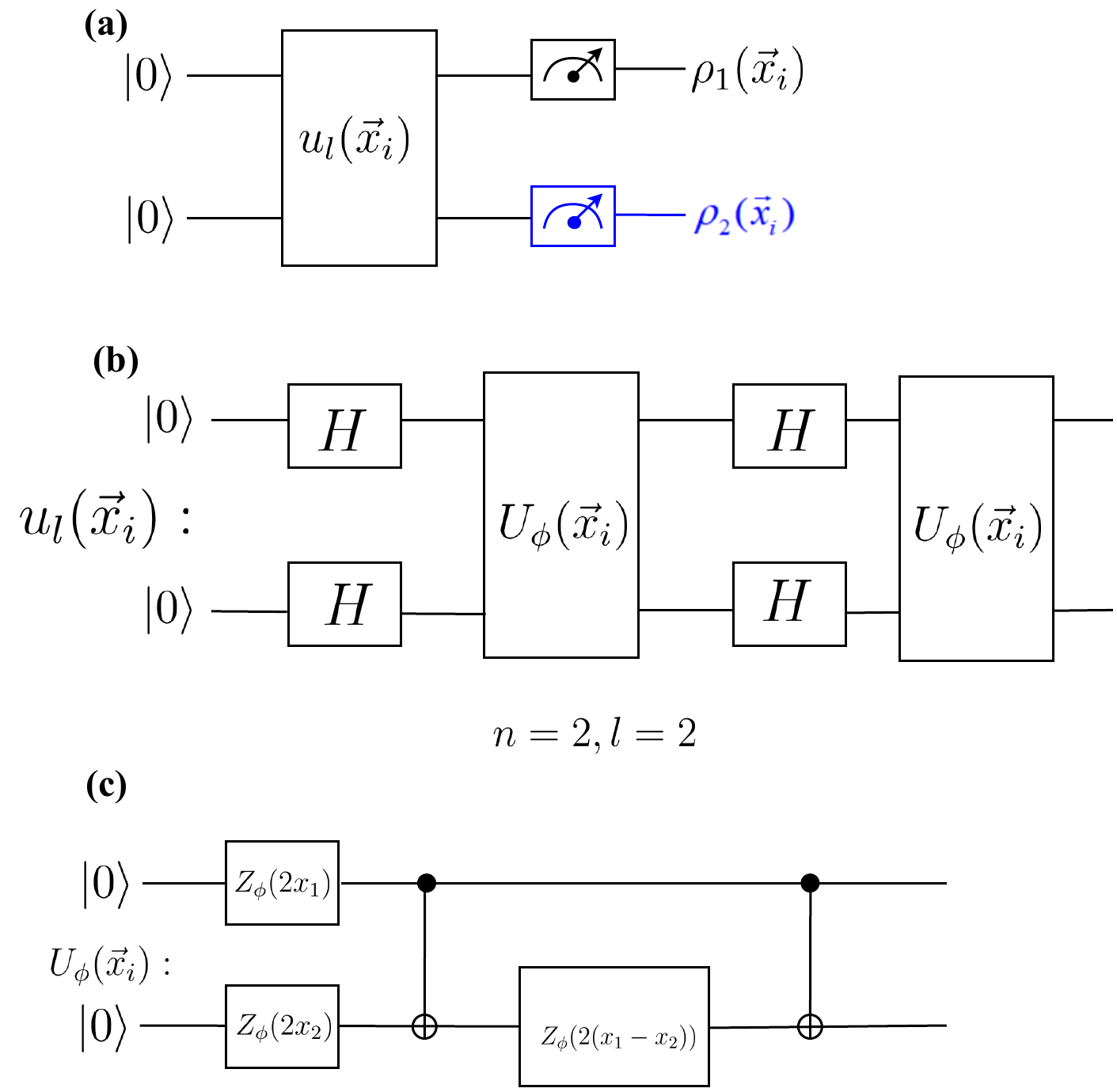}
\caption{\textbf{(a)}~A schematic picture of a two-qubit version of the projected kernel circuit. For the projected quantum kernel all qubits are measured, and moreover, all inputs are pure states. Here, for a two-qubit version, measuring the first qubit reconstructs the subsystem $\rho_1(\vec x_i)$, and then measuring the second qubit (indicated in blue)  reconstructs $\rho_2(\vec x_i)$. \textbf{(b)}~The gate decomposition for the unitary matrix ${u^{l}(\vec x_i)}$, where $l$ is the number of the iterations of this gate decomposition (length of the circuit) for two qubits $n=2$, and two iterations $l=2$. \textbf{(c)}~ The gate components of the unitary operator $U_\phi{(\vec{x_i})}$ adapted from~\cite{havlivcek2019supervised}. Here, $H$ is the Hadamard gate, and $Z_\phi=\text{diag}(1,e^{i\phi(\vec{x_i})})$ is a single qubit phase gate.}
\label{pkernel}
\end{figure}

It is worth noting the similarity between our approach and the projected kernel method introduced in~\cite{Huang_2021}, where both methods involve constructing the kernel function by measuring a subset of the relevant qubits. More explicitly, the projected kernel function is defined as~\cite{Huang_2021} 
\begin{align}
k^{PQ}(\vec x_i,\vec x_j) = \exp \left( -\gamma \sum_{m=1}^n \norm{\rho_m(\vec x_i) - \rho_m(\vec x_j)}^2\right),
\end{align}
where $\rho_m(\vec x_i)$ is the reduced density matrix of the $m$-th register of the encoded quantum state $\rho(\vec x_i)$, and $\gamma>0$ is a hyperparameter. 
Furthermore, their work extends this kernel to $k^{(PQ)}_s$ which takes every subset of $s$ qubits into account (for clarification, we have $k^{(PQ)}_1=k^{(PQ)}$). We highlight that
the number of measurements required in determining $k^{(PQ)}_s$ grows as $4^s {n\choose s}$ since there are $4^s$ Pauli strings on $s$ qubits, and ${n\choose s}$ subsets of size $s$ for a state of $n$ qubits. This quantity grows polynomially in $n$ and exponentially in $s$. 
In contrast, our method only necessitates the measurement of the control qubit. This practical aspect holds strong appeal because, in current superconducting qubit technology, measurements are the most error-prone operations, with error rates ranging from 3 to 10 times that of 2-qubit gates, as demonstrated in~\cite{arute2019quantum}.
For completeness, we repeated our experiments employing the projected quantum kernel on ``ad\_hoc'' dataset~\cite{code}. The two-qubit projected kernel is depicted in Fig. \ref{pkernel}a with the same encoding circuit (Fig. \ref{pkernel}b), and feature map (Fig. \ref{pkernel}c) as in DQC1 kernel. The Qiskit simulation and IBM hardware's results (considering $\gamma=0.01$) demonstrated an accuracy of $100\%$ and $90\%$, respectively.

\section{THE ROLE OF COHERENCE AND THE EFFECT OF NOISE}\label{role of coh}

In the following, we explore the role of control qubit's coherence, hardware noise, coherence consumption and quantum discord production in our setting.  

\begin{figure}
\centering
\includegraphics[width=0.4\textwidth]{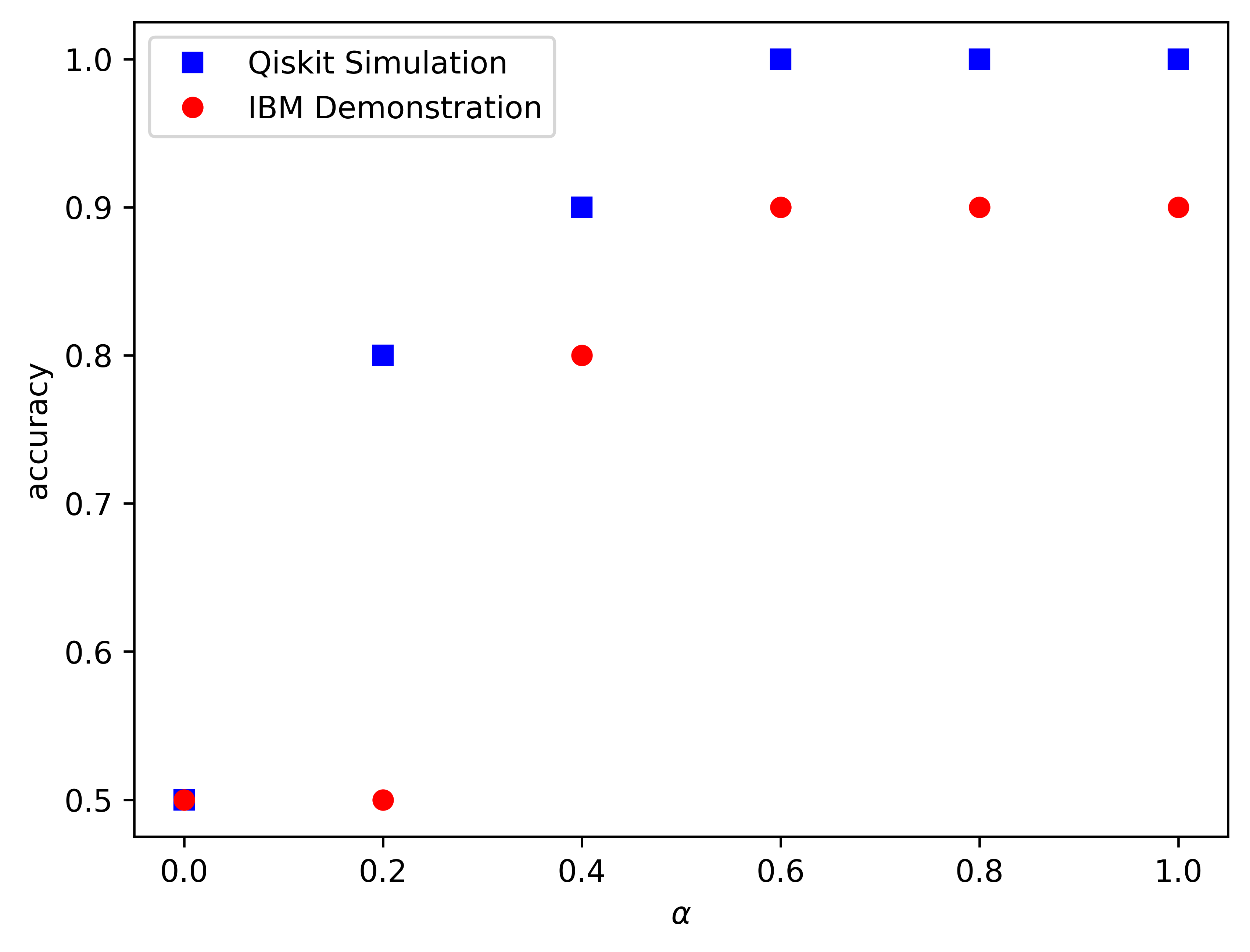}
\caption{ The accuracy as a function of the control qubit's purity for the same dataset as in Fig.(\ref{adhoc-classification}) is shown. Note that when $\alpha=0$, the state is in a completely mixed state, and when $\alpha=1$, the state is pure. The blue curve indicates simulation results, and the red curve shows the results obtained from IBM hardware.}
\label{alpha}
\end{figure}

To see the role of the control qubit's coherence in our implementation, we repeat the learning task  with the control qubit prepared in the state $\frac{I_1+\alpha Z}{2}$, where $0\leq\alpha\leq1$.
In Fig. \ref{alpha} we show the prediction accuracy in the simulation (blue dots) and the implementation (red dots) as a function of the purity of the control qubit. From Fig. \ref{alpha} one can see that $\alpha=0$, for which the control qubit is in a maximally mixed state, the accuracy is $0.5$, corresponding to randomly guessing the labels. By increasing the purity, however, the accuracy increases until it reaches its maximum value at $\alpha\geq 0.6$. Due to the device noise, the accuracy in the implementation is degraded in comparison to the simulation.

\begin{figure}
\centering
\includegraphics[width=0.4\textwidth]{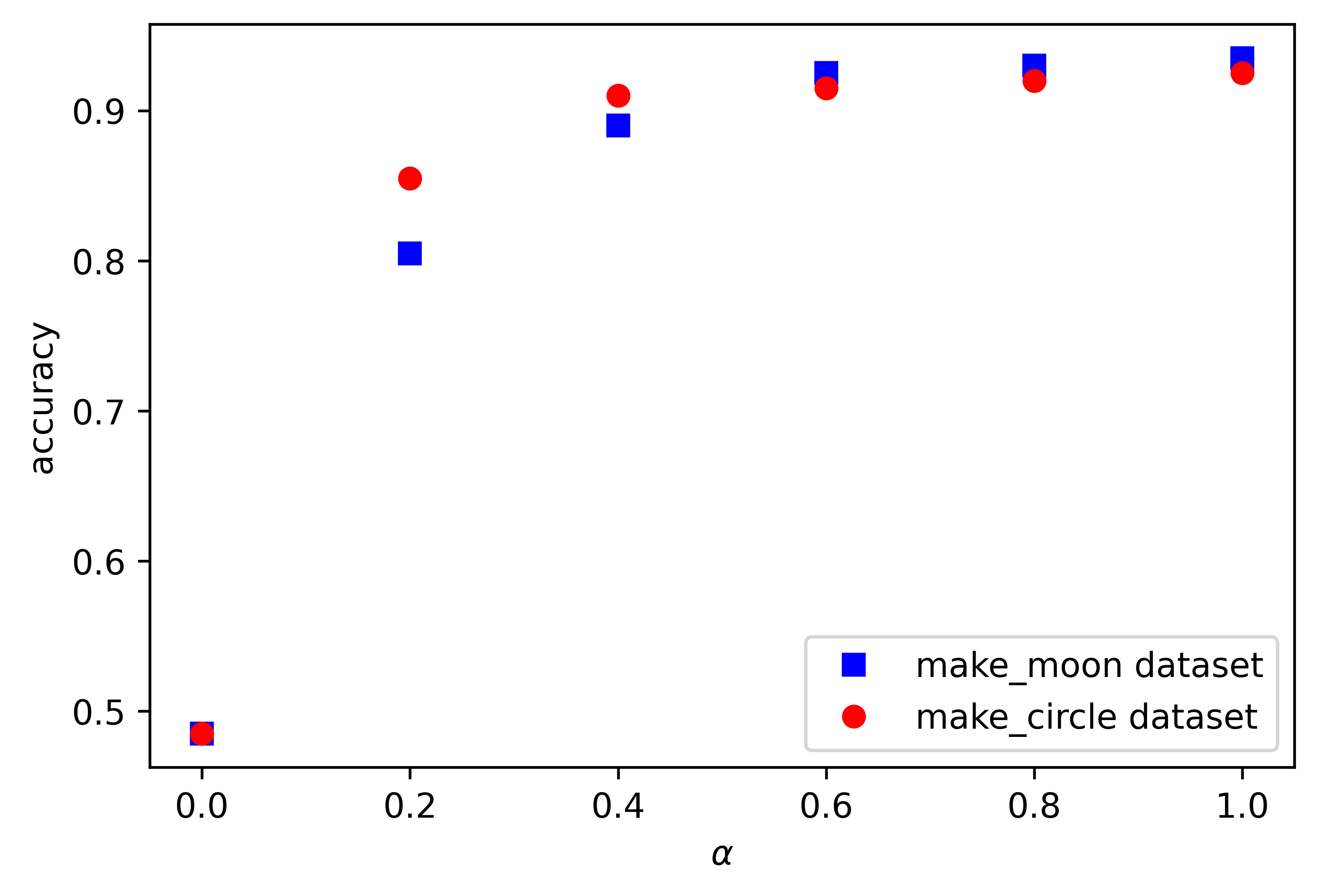}
\caption{Simulation results of learning process for ``make-moon'' (blue curve), and ``make-circle'' dataset (red curve). The maximum accuracies, being $0.935$, and $0.925$, are both achieved at $\alpha=1$.}
\label{learning-different datapoints}
\end{figure}

For completeness we repeat the learning process with two well-known datasets called ``make-moon'', and ``make-circle'' from ``scikit-learn'', each of them including $800$ training data points, and $200$ testing data points. For these two datasets we observed an abrupt change in the accuracy for $\alpha\geq0.2$. Hence the critical value of $\alpha$ depends on the dataset. These results are depicted in Fig.~\ref{learning-different datapoints}. By interpreting $1-\alpha$ as the noise strength, one can see that the accuracy is robust against noise for $\alpha\geq0.6$ (Fig. \ref{alpha}) and $\alpha\geq0.2$ (Fig. \ref{learning-different datapoints}). Likewise, variational quantum circuits are predicted to display similar robustness against noise~\cite{liu2022laziness}. Let us emphasize that in the rest of our paper we use $\alpha=1$.

\begin{figure}
\centering
\includegraphics[width=0.5\textwidth]{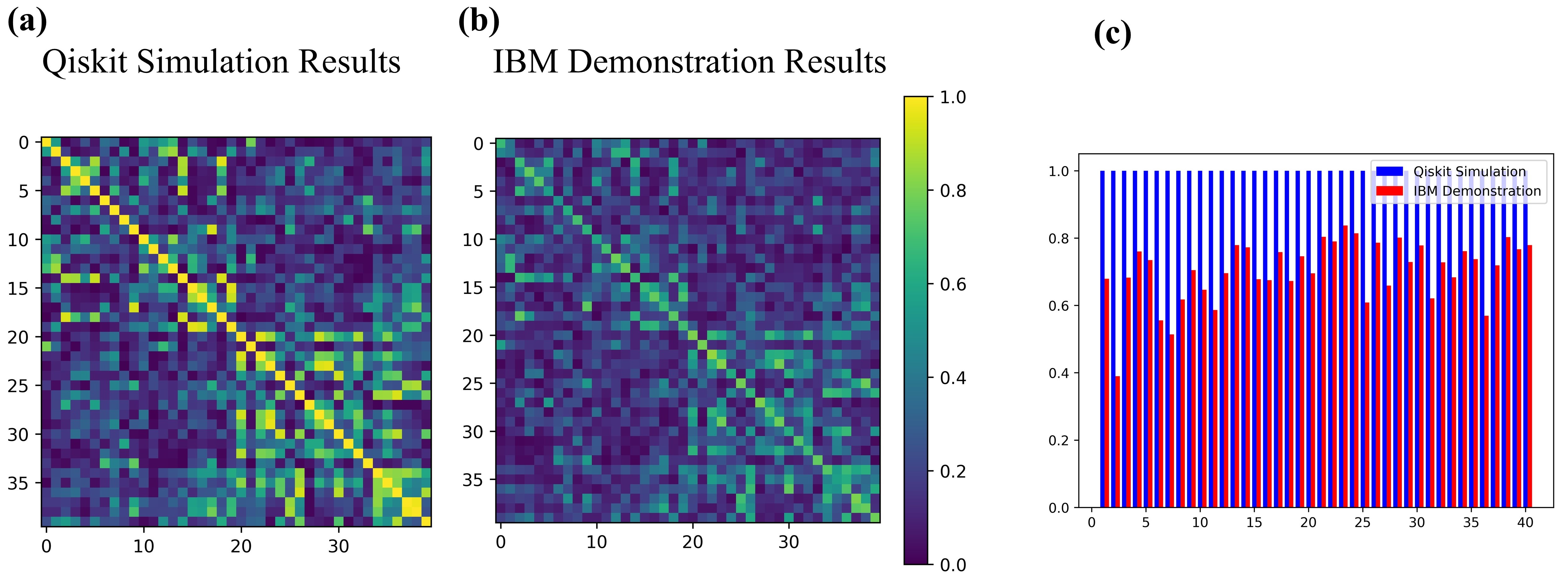}
\caption{\textbf{(a)}. The Qiskit simulation (left), and \textbf{(b)}. IBM (right) results for the DQC1 quantum kernel with $n=2$, and $l=2$. IBM results have been obtained from ``$ibm\_perth$'' device.
\textbf{(c)}. Diagonal elements of simulated (blue bars), and IBM (red bars) kernel matrices. The maximum difference between diagonal elements is $0.610$, and the mean difference is $0.27$.}
\label{qkernels}
\end{figure}

\begin{figure}
\centering
\includegraphics[width=0.4\textwidth]{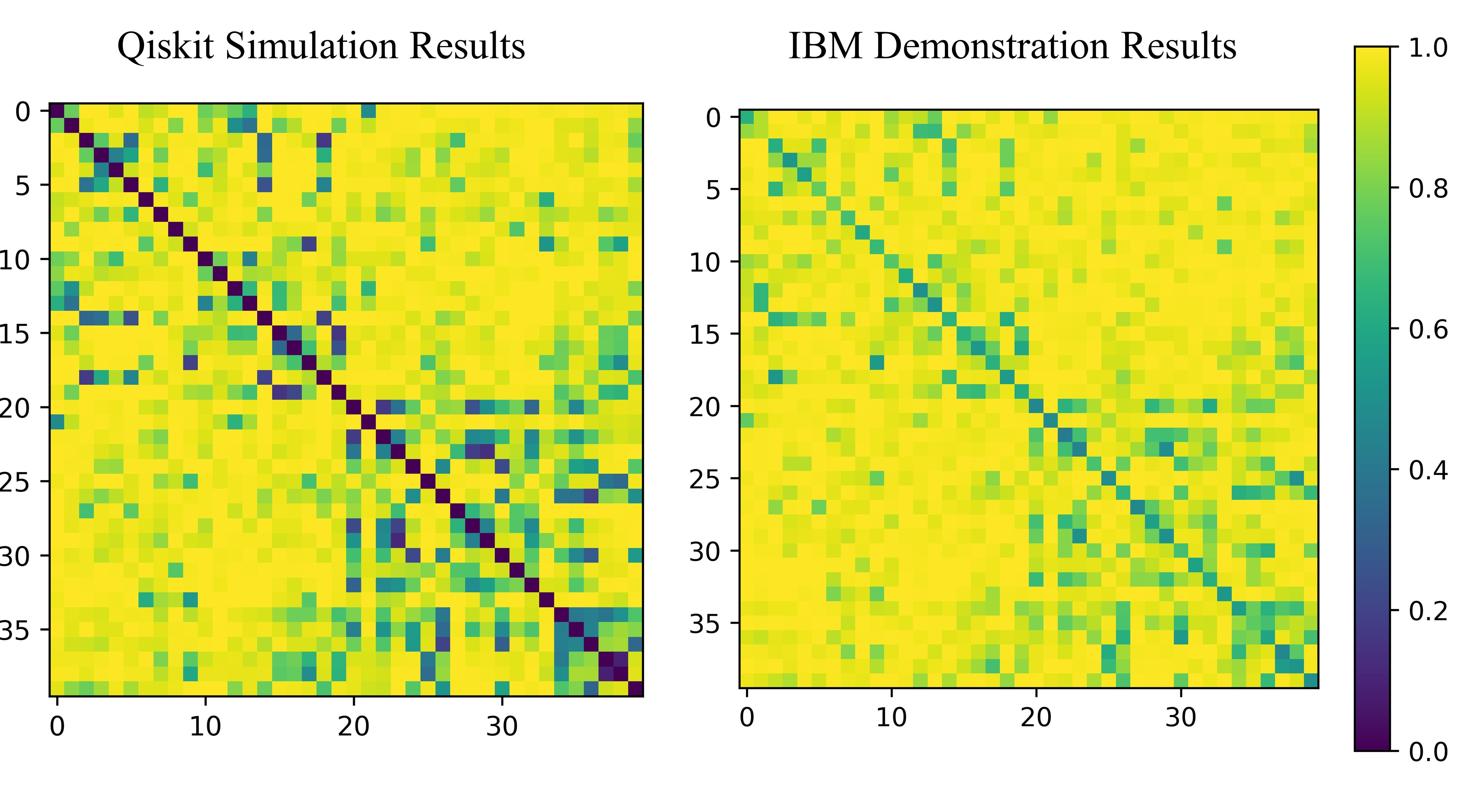}
\caption{The Qiskit simulation results (left), and IBM results (right) for coherence consumption for the same dataset as in Fig.~\ref{adhoc-classification}, and the circuit in Fig.~\ref{dqc}.}
\label{coherence}
\end{figure}

In Fig.\ref{qkernels}(a,b) we show the absolute value of the kernel obtained from simulation, and IBM hardware respectively. The difference between the two kernels can be attributed to hardware noise. To better show the role of noise in the kernel, we compare the diagonal elements of the kernel obtained from simulation and IBM hardware. As discussed earlier, in the ideal case $K(\vec x,\vec x) = 1$ (blue bar) but in practice one has $K(\vec x,\vec x)<1$. In Fig.\ref{qkernels}(c), one finds a maximum difference of $0.610$ between simulation and implementation, while the mean difference is $0.27$.
Having access to the kernel, we can obtain the coherence consumption in our implementation from Eq. (\ref{deltac2}), as shown in Fig.\ref{coherence}. In accordance with Eq.(\ref{deltac2}), it can be seen from Fig.\ref{coherence} that the coherence consumption is minimum (but not equal to zero in the IBM results) along the diagonal axes.

\begin{figure}
\centering
\includegraphics[width=0.4\textwidth]{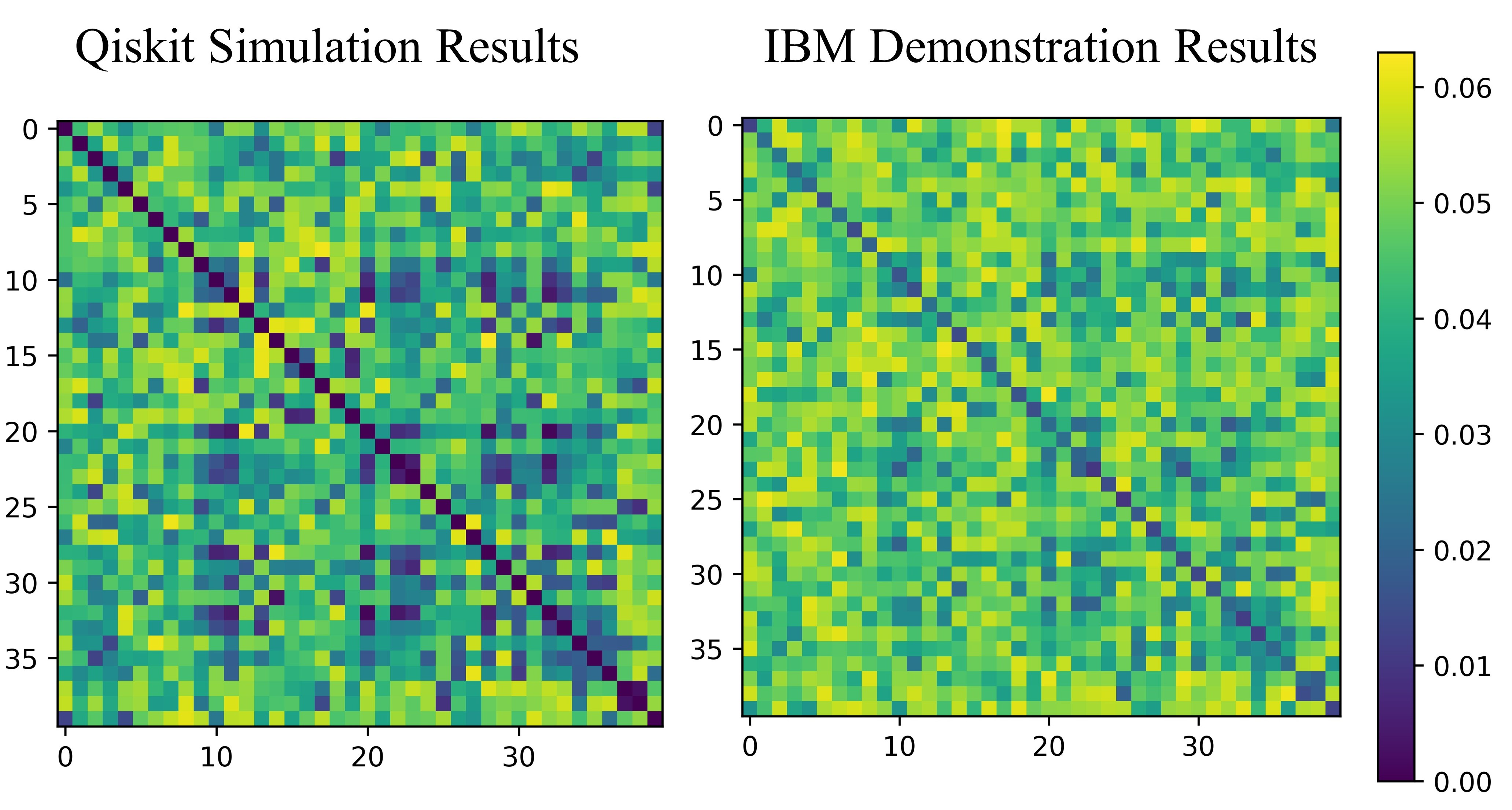}
\caption{The Qiskit simulation results (left), and IBM results (right) for geometric discord for the same dataset as in Fig.~\ref{adhoc-classification}, and the circuit in Fig.~\ref{dqc}.}
\label{gdiscord}
\end{figure}

Our next step is to obtain the generated discord in our implementation based on Eq.(\ref{disdqc1}). Eq.(\ref{disdqc1}) indicates that for estimating the discord, $\text{tr}(U_{n}^{2})$ must be estimated, which requires successive application of DQC1 evolution Eq.(\ref{Udqc1}). Fig.\ref{gdiscord} shows the quantum discord. When comparing Fig.\ref{gdiscord} with Fig.\ref{coherence}, it is also evident that the condition (\ref{disbound}) is satisfied.

\section{DISCUSSION \& CONCLUSION}\label{conclusion} 

 In \cite{heyraud2022noisy}, an upper bound on the generalization error of the fidelity quantum kernel model has been established, which is determined by the average purity of the encoded states. As highlighted in the same work, a noisier encoding process, when transferring data into quantum states, can lead to poorer training performance. Our empirical results, illustrated in Figures~\ref{alpha} and~\ref{learning-different datapoints} and based on three distinct datasets, support the idea that improving the purity of the control qubit (parameterized by $\alpha$) enhances the accuracy of DQC1 kernel model.  We believe that our DQC1 kernel framework provides a ground for rigorous theoretical studies for relating coherence consumption to generalization error.

We also comment on the connection between our work and the findings in \cite{thanasilp2022exponential}, which investigate the exponential concentration of quantum kernels. Despite our method relying solely on measuring the control qubit, it is essential to emphasize that the number of required measurements scales exponentially with the number of target qubits, as the variance in the kernel exponentially approaches zero. Hence, we expect the same untrainability issues to hold for our kernel function.\\

In this study, we have investigated the application of the DQC1 model, a restricted computational model, to supervised machine learning tasks. Unlike the standard universal computational model, the DQC1 model relies on mixed states and does not incorporate quantum entanglement into the computation. We have presented a test of the DQC1 model's ability to solve supervised machine learning problems for some classically difficult kernels~\cite{havlivcek2019supervised}. Despite requiring a greater number of gates than a similar protocol described in~\cite{havlivcek2019supervised}, since one needs to measure only the control qubit, our protocol still achieved a relatively high level of classification accuracy. Our proposal highlights the potential of utilizing quantum discord over entanglement in the presence of noise.

In a broader context, our work highlights the computational power of a single-qubit as a universal classifier~\cite{perez2020data,dutta2022single}. %There is limited literature on the application of DQC1 in machine learning~\cite{park2018noise,ghobadi2019power}. 
It would be interesting to realize our protocol in the NMR setting~\cite{kusumoto2021experimental}. We hope that this study will inspire further research on the integration of quantum coherence and quantum discord in machine learning.

\section{Acknowledgement} 
We acknowledge discussions with Anton Dekusar and Hadi Zadeh-Haghighi. This work was supported by the Alberta Major Innovation Fund (MIF) Quantum Technologies project and by the Natural Sciences and Engineering Research Council (NSERC) of Canada.

\section{Author contributions}
M.K. performed simulation analysis. A.J-A. wrote the optimization code to run the code on IBM hardware. C.S. provided detailed feedback on the manuscript. R.G. supervised the entire project and wrote the manuscript with input from M.K., A.J-A. and C.S.

\bibliographystyle{apsrev4-1}
\bibliography{main.bib}

\onecolumngrid

\appendix

\section{Appendix: Deriving Eq.(\ref{deltac})}

The input state of the control qubit can be obtained by tracing out the Eq.(\ref{rhoin}) 

\begin{align}\label{rhoinc}
\rho_{in,c} =\frac{1}{2}
\begin{pmatrix}
 1& \alpha\\
\alpha & 1
\end{pmatrix}
= \frac{1+\alpha}{2} \ket{+}\bra{+} + \frac{1-\alpha}{2} \ket{-}\bra{-},\tag{$A.1$}
\end{align}
where $\ket{+}=\frac{\ket{0}+\ket{1}}{\sqrt{2}}$ and $\ket{-}=\frac{\ket{0}-\ket{1}}{\sqrt{2}}$.
Using the definition of von Neumann entropy $[S(\rho)=-\text{tr}(\rho\text{log}\rho)]$ for Eq.(\ref{rhoinc}) it is straightforward to obtain $S(\rho_{diag,in,c}) = 1$ and $S(\rho_{in,c}) = H_{2}\left(\frac{1-\alpha}{2}\right)$, where we used the binary entropy $H_{2}(x)=-x\text{log}_{2}x-(1-x)\text{log}_{2}(1-x)$. Therefore upon using Eq.(\ref{cohdef}) one obtains

\begin{align}\label{ci}
C(\rho_{in,c}) = 1-H_{2}\left(\frac{1-\alpha}{2}\right).\tag{$A.2$}
\end{align}

Next, we obtain the coherence in the final state Eq.(\ref{rhof}). Noting that the eigenvalues of Eq.(\ref{rhof}) are $\mu_\pm = \frac{1\pm \abs{\alpha\tr U}/2^n}{2}$, we get $S(\rho_{f,c}) = H_2\left(\frac{1-\abs{\alpha \tr U}/2^n}{2}\right)$. Also, since all the diagonal entries of $\rho_{f,c}$ are equal, we have $S(\rho_{diag, f,c})=1$. Thus, we obtain

\begin{equation}\label{cf}
C(\rho_{f,c})=1- H_2 
 \left(  
 \frac{1- \abs{\alpha \tr U}/2^n}{2}
 \right).\tag{$A.3$}
\end{equation}

Finally, $\Delta C=C(\rho_{in,c})-C(\rho_{f,c}) $ together with Eqs. (\ref{ci},\ref{cf}) conclude
\begin{equation}
\Delta C =H_2 
 \left(  
 \frac{1- \abs{\alpha \tr U}/2^n}{2}
 \right) - H_{2}\left(\frac{1 -\alpha}{2}\right),\tag{$A.4$}
\end{equation}

which is the Eq.(\ref{deltac}).

\end{document}